\documentclass[aps,prb,twocolumn,superscriptaddress,showpacs]{revtex4}
\usepackage{amsmath}
\usepackage{amssymb}
\usepackage{url}
\usepackage{graphicx}
\usepackage{dcolumn} 
\usepackage{bm}      
\usepackage{multirow}
\usepackage{epstopdf}
\usepackage{color}
\usepackage{gensymb}
\usepackage{textcomp}

\begin{document}

\title{Canted Magnetic Ground State of Quarter-Doped \\
Manganites $R_{0.75}$Ca$_{0.25}$MnO$_3$ ($R$ = Y, Tb, Dy, Ho, and Er)}

\author{R.~Sinclair}
\affiliation{Department of Physics and Astronomy, University of Tennessee, Knoxville, Tennessee 37996-1200, USA}

\author{H.~B.~Cao}
\affiliation{Quantum Condensed Matter Division, Oak Ridge National Laboratory, Oak Ridge, Tennessee, 37831, USA}

\author{V.~O.~Garlea}
\affiliation{Quantum Condensed Matter Division, Oak Ridge National Laboratory, Oak Ridge, Tennessee, 37831, USA}

\author{M.~Lee}
\affiliation{Department of Physics, Florida State University, Tallahassee, Florida 32306, USA}
\affiliation{National High Magnetic Field Laboratory, Florida State University, Tallahassee, Florida 32310, USA}

\author{E.~S.~Choi}
\affiliation{National High Magnetic
Field Laboratory, Florida State University, Tallahassee, Florida 32310, USA}

\author{Z.~L.~Dun}
\affiliation{Department of Physics and Astronomy, University of Tennessee, Knoxville, Tennessee 37996-1200, USA}

\author{S.~Dong}
\email{sdong@seu.edu.cn}
\affiliation{Department of Physics, Southeast University, Nanjing 211189, China}

\author{E.~Dagotto}
\affiliation{Department of Physics and Astronomy, University of Tennessee, Knoxville, Tennessee 37996-1200, USA}
\affiliation{Materials Science and Technology Division, Oak Ridge National Laboratory, Oak Ridge, Tennessee 37831, USA}

\author{H.~D.~Zhou}
\email{hzhou10@utk.edu}
\affiliation{Department of Physics and Astronomy, University of Tennessee, Knoxville, Tennessee 37996-1200, USA}
\affiliation{National High Magnetic Field Laboratory, Florida State University, Tallahassee, Florida 32310, USA}
\date{\today}

\begin{abstract}
Polycrystalline samples of the quarter-doped manganites $R_{0.75}$Ca$_{0.25}$MnO$_3$ ($R$ = Y, Tb, Dy, Ho, and Er) were studied by X-ray diffraction and AC/DC susceptibility measurements. All five samples are orthorhombic and exhibit similar magnetic properties: enhanced ferromagnetism below $T_1$ ($\sim80$ K) and a spin glass (SG) state below $T_{SG}$ ($\sim30$ K). With increasing $R^{3+}$ ionic size, both $T_1$ and $T_{SG}$ generally increase. The single crystal neutron diffraction results on Tb$_{0.75}$Ca$_{0.25}$MnO$_3$ revealed that the SG state is mainly composed of a short-range ordered version of a novel canted (i.e. noncollinear) antiferromagnetic spin state. Furthermore, calculations based on the double exchange model for quarter-doped manganites reveal that this new magnetic phase provides a transition state between the ferromagnetic state and the theoretically predicted spin-orthogonal stripe phase.
\end{abstract}

\maketitle

\section{INTRODUCTION}
Recent research in the field of manganites has primarily
focused on the multiferroic properties
of $R$MnO$_3$.\cite{kimura, goto, kenzelmann, kimura2, arima, mostovoy, lorenz, hur, chapon} This includes studies of Type-I multiferroics employing materials such as the hexagonal YMnO$_3$ where ferroelectricity and magnetism have different origins,\cite{Aken:Nm} and also
studies of Type-II multiferroics as in the cases of TbMnO$_3$ and orthorhombic
HoMnO$_3$ where the ferroelectricity is caused by peculiar  magnetic
orders.\cite{Sergienko:Prb,Sergienko:Prl,Picozzi} All of these multiferroic $R$MnO$_3$ materials
are located in the narrow-bandwidth limit of manganites due to their small $R^{3+}$
ionic size.\cite{SizeMismatchPRB} The fact that in Type-II multiferroics
the ferroelectricity is strongly coupled to magnetism makes it very exciting
to explore the possibility of new magnetic states in narrow-bandwidth systems.
Discovering exotic magnetic phases may lead to functional multiferroics
with high transition temperatures and large spontaneous polarizations.

While the cases of pure undoped manganites have been extensively studied,\cite{Ishiwata:Prb10,Dong:Mplb}
the recent exploration for potential new magnetic phases has focused on
doped manganites in the narrow-bandwidth limit.\cite{RCaJPhys,LnCaJSSC}
The competition between the ferromagnetic double exchange interactions
and the antiferromagnetic superexchange interactions, plus the robust
Jahn-Teller spin-lattice coupling, makes the doped narrow-bandwidth manganites
an ideal playground to search for new magnetic phases.\cite{MagnetsCCR} Moreover, considerable
theoretical progress has been made in this area of research. For example,
the magnetic phase diagram of quarter-doped manganites has been investigated based
on the double exchange model.\cite{SOSQdopedPRL} A prominent ferromagnetic (FM)
metallic orbitally-disordered phase occupies the region where the superexchange interaction
between the $t_{2g}$ spins, $J_{\rm AFM}$, is small in agreement
with the phase diagram of large and intermediate bandwidth manganites. However, a new exotic multiferroic
phase, dubbed the spin-orthogonal-stripe (SOS) phase, was found at the large $J_{\rm AFM}$
end of the narrow-bandwidth manganites. The SOS state, as shown in Fig. 2(a) of Ref.~\onlinecite{SOSQdopedPRL},
is made of zigzag chains, as in the CE-phase of manganites at quarter doping,
but forming an array of diagonally oriented domains with spins rotated by $90\degree$  between domains.
Generalizations to other hole dopings have been studied as well.\cite{SOSgen}

The intriguing prediction of an SOS state requires detailed
experimental studies. Although a couple of previous efforts have been reported
for the doped manganites with small  $R^{3+}$ ions, such as
Tb$_{1-x}$Ca$_x$MnO$_3$,\cite{TbCaPRB1, TbCaPRB2, TbCaJPhys, TbCaPhysJap1, TbCaPhysJap2, TbCaPB}
Dy$_{1-x}$Ca$_x$MnO$_3$,\cite{DyCaSSS} and Y$_{1-x}$Ca$_x$MnO$_3$,\cite{YCaSSC, YCaAPA}
a systematic exploration of quarter-doped manganites $R_{0.75}$Ca$_{0.25}$MnO$_3$ has not
been conducted and, for this reason, the existence of the SOS phase, or other exotic magnetic phases, has not been confirmed so far.
Moreover, theoretically the nature of the quarter-doped manganites
with an intermediate $J_{\rm AFM}$ strength
has not been studied in full detail: it is only known that the FM and SOS states
are stable in the limits of small and large $J_{\rm AFM}$, respectively.\cite{SOSQdopedPRL}
Thus, an intriguing question develops: does any new magnetic phase exist in the intermediate coupling region in between the FM and the predicted SOS phases?
Plenty of previous theoretical studies have consistently
shown that manganites in general, i.e. not only multiferroics, have the potential
to display a wide variety of complex patterns of spin, charge, and orbital order.\cite{Hotta99,Hotta03,Hotta01,Hotta00,Dong08}

With this timely fresh motivation in mind to search for possible new magnetic phases
of quarter-doped manganites with a narrow-bandwidth, in this publication we have
systematically studied the magnetic characteristics of $R_{0.75}$Ca$_{0.25}$MnO$_3$
($R$ = Tb, Dy, Ho, Y, and Er). First, we analyzed the magnetic properties
of all five polycrystalline samples and found that their general behavior
consists of enhanced FM interactions below $80$ K and a spin glass state
below $30$ K. Second, we performed neutron scattering experiments
on a single crystal of Tb$_{0.75}$Ca$_{0.25}$MnO$_3$ and discovered
that its magnetic ground state is dominated by the short range ordering
of a new canted spin state which results in the observed
spin glass-like behavior. Finally, we conducted theoretical calculations
based on the double exchange model confirming that the observed
new spin structure is lower in energy than both the FM and SOS states
in the intermediate strength region of $J_{\rm AFM}$; in other words,
the novel magnetic phase reported here acts as a transition state between the FM and the SOS phases.

\section{EXPERIMENTAL DETAILS}

Polycrystalline samples of $R_{0.75}$Ca$_{0.25}$MnO$_3$ ($R$ = Tb, Dy, Ho, Y, and Er) were synthesized by solid state reactions. The stoichiometric mixture of
Tb$_4$O$_7$/Dy$_2$O$_3$/Ho$_2$O$_3$/Y$_2$O$_3$/Er$_2$O$_3$, CaCO$_3$,
and Mn$_2$O$_3$ were ground together and then calcined
in air at 950 $^{\circ}$C,  1200 $^{\circ}$C, and 1350 $^{\circ}$C for 24 hours,
respectively. Single crystals of Tb$_{0.75}$Ca$_{0.25}$MnO$_3$ were grown by the
traveling-solvent floating-zone (TSFZ) technique in an IR-heated image furnace (NEC)
equipped with two halogen lamps and double ellipsoidal mirrors. The crystal growth
rate was 15 mm/h. Small pieces of single crystals were ground into a
fine powder for X-ray diffraction. The resulting powder X-ray diffraction (XRD) patterns
were recorded at room temperature with a HUBER Imaging Plate Guinier Camera 670
with Ge monochromatized Cu $K_{\alpha1}$ radiation (1.54059 {\AA}). The lattice
parameters were refined from the XRD patterns by using the software package \textit{FullProf Suite} with typical refinements for all samples having $\chi^2\approx0.7$. X-ray Laue diffraction was used to align the crystals. Elastic neutron scattering measurements were performed at the Neutron Powder Diffractometer (HB-2A), and single-crystal neutron scattering measurements were performed at the Four-Circle Diffractometer (HB-3A). Both instruments are located at the High Flux Isotope Reactor (HFIR) in Oak Ridge National Laboratory (ORNL). The neutron scattering diffraction patterns were also refined using \textit{FullProf Suite}. The DC magnetic-susceptibility measurements were performed employing a Quantum Design superconducting interference device (SQUID) magnetometer. The AC susceptibility data was measured on a homemade setup.\cite{ZLD}

\section{RESULTS}

\subsection{Polycrystalline $R_{0.75}$Ca$_{0.25}$MnO$_3$}

\begin{figure*}[tbp]
\linespread{1}
\par
\begin{center}
\includegraphics[width=\textwidth]{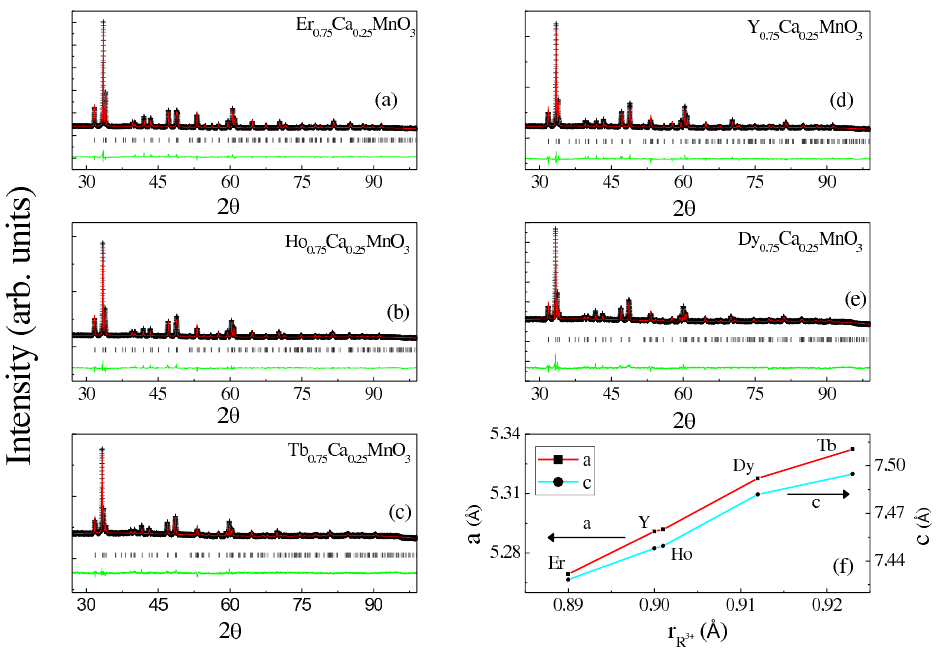}
\end{center}
\par
\caption{(color online) Room temperature XRD patterns for polycrystalline
$R_{0.75}$Ca$_{0.25}$MnO$_3$: (a) Er, (b) Ho, (c) Tb, (d) Y, and (e) Dy.
The crosses are the experimental data. The solid curves are the best fits
from the Rietveld refinement using \textit{FullProf Suite}. The vertical marks
indicate the position of the Bragg peaks, and the bottom curves show
the difference between the observed and calculated intensities. (f)
The $R^{3+}$ ionic size dependence of the lattice parameters.}
\end{figure*}

Figures 1 (a-e) display the room temperature XRD patterns for
$R_{0.75}$Ca$_{0.25}$MnO$_3$ ($R$ = Tb, Dy, Ho, Y, and Er). All samples show a
pure orthorhombic ($Pbnm$) structure. With decreasing ionic size $R$, the
lattice parameters decrease, as shown in Fig.~1(f). The attempts to prepare
orthorhombic $R_{0.75}$Ca$_{0.25}$MnO$_3$ with $R^{3+}$ ions smaller than Er$^{3+}$
failed. Noting that pure HoMnO$_3$, YMnO$_3$, and ErMnO$_3$ have hexagonal structures,
it seems that the substitution of Ca$^{2+}$ can stabilize the perovskite structure.
This is easy to understand since Ca$^{2+}$ ions are larger than the $R^{3+}$ ions involved
here and this doping will reduce the manganese size. Therefore, the structural tolerance
factor ($t$) will be increased and, generally, as $t$ increases the driving force
for the octahedral rotation increases as well,  leading to the transformation
from the hexagonal to the orthorhombic phase. Several reported studies on
Ho$_{1-x}$Ca$_x$MnO$_3$ and Y$_{1-x}$Ca$_x$MnO$_3$ have
confirmed this observation.\cite{RCaJPhys, YCaAPA, HoCaJMaterSci}

\begin{figure*}[tbp]
\linespread{1}
\par
\begin{center}
\includegraphics[width=5in]{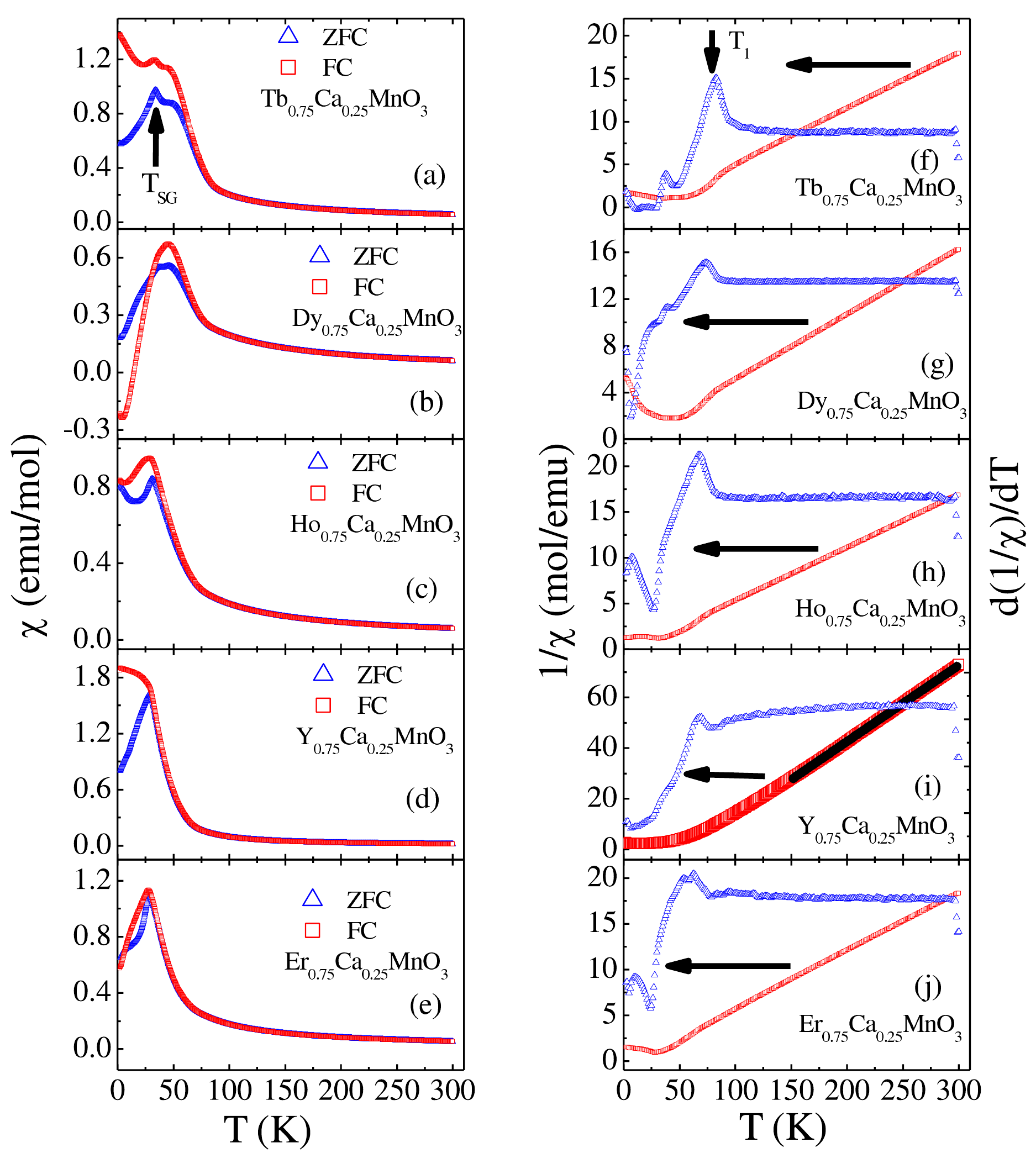}
\end{center}
\par
\caption{(color online) Temperature dependence of the DC susceptibility
for polycrystalline $R_{0.75}$Ca$_{0.25}$MnO$_3$: (a) Tb, (b) Dy, (c) Ho,
(d) Y, and (e) Er.  Temperature dependence of the inverse of the susceptibility
and its derivative for (f) Tb, (g) Dy, (h) Ho, (i) Y, and (j) Er. In (i),
the linear solid line represents the Curie-Weiss fit for the $T$ $>$ 150 K data.}
\end{figure*}

\begin{figure}[tbp]
\linespread{1}
\par
\begin{center}
\includegraphics[width=3.6in]{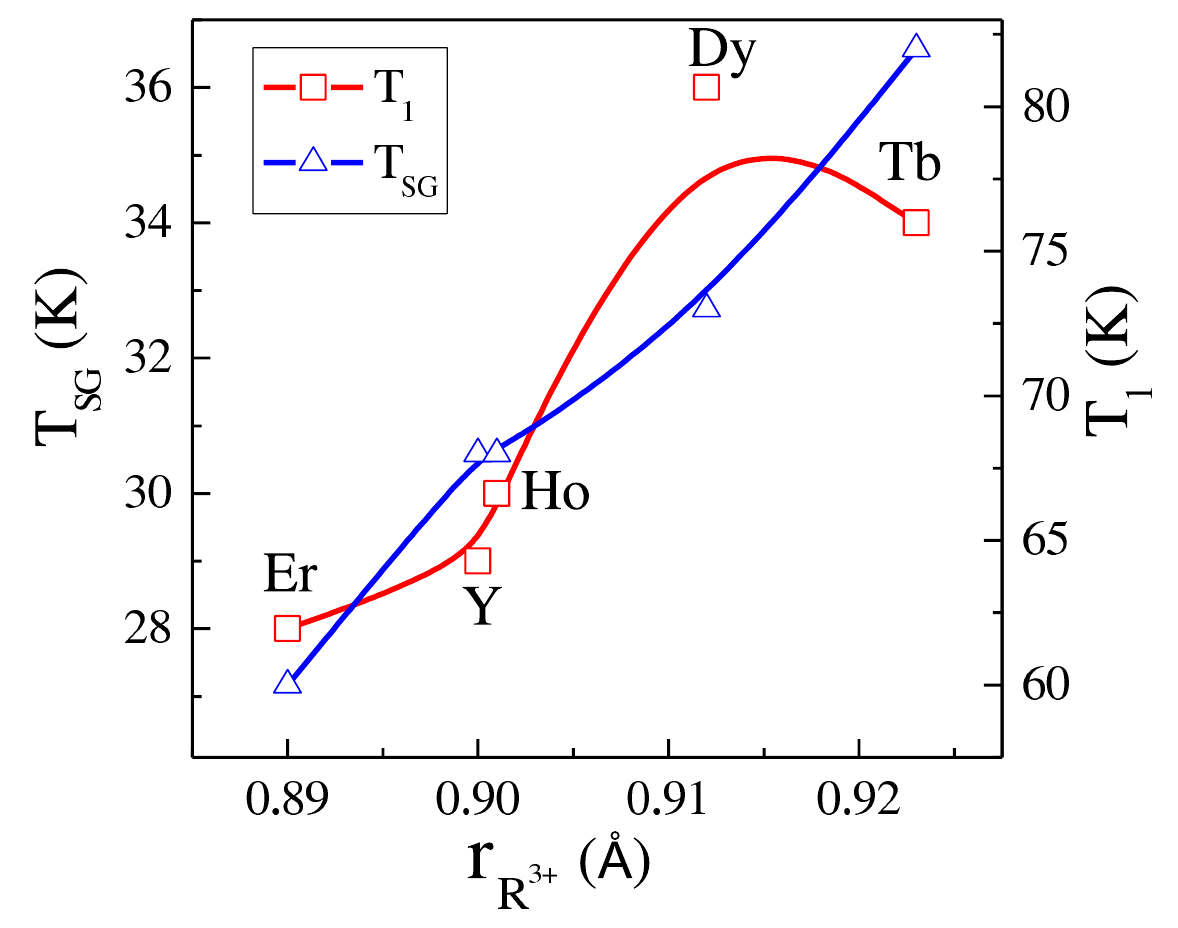}
\end{center}
\par
\caption{(color online) Variations of the $T_1$ and $T_{SG}$ temperatures
with the $R^{3+}$ ionic size for the $R_{0.75}$Ca$_{0.25}$MnO$_3$ materials studied here.}
\end{figure}

Figure 2 shows the DC susceptibility ($\chi$) results corresponding to $R_{0.75}$Ca$_{0.25}$MnO$_3$.
For the case of Tb$_{0.75}$Ca$_{0.25}$MnO$_3$, $\chi$ shows two major features:
(i) a sharp increase around 80 K with decreasing temperature.
This feature is more evident as a slope change from the linear temperature dependence
of the inverse of the susceptibility (Fig. 2(f)). Here we define the peak position of
the derivative of 1/$\chi$ as $T_1$. Clearly, ferromagnetic tendencies develop below $T_1$; (ii) a broad peak around 34 K where zero field cooling (ZFC) and field cooling (FC)
curves display a large splitting. This peak position is defined as $T_{SG}$.
Similarly, all other samples also present these two features as well. As shown in Fig. 3,
with increasing $R^{3+}$ ionic size, both $T_1$ and $T_{SG}$ generally increase.
One important point here is that since Y$_{0.75}$Ca$_{0.25}$MnO$_3$ with nonmagnetic
Y$^{3+}$ ions display these two features, they must be related to the
magnetic Mn$^{3+}$/Mn$^{4+}$ ions. To further investigate the magnetic properties,
we performed a Curie-Weiss fit
on the 1/$\chi$ data, as shown in Fig. 2(i), resulting in a Curie temperature of
$\theta_{CW}$ = 58.1 K and an effective magnetic moment
of $\mu_{eff}$ = 5.16 $ $ $\mu_{B}$.
For Y$_{0.75}$Ca$_{0.25}$MnO$_3$, there are 75{\%} Mn$^{3+}$ ions
($\mu_{eff}$ $\approx$ 4.8 $ $ $\mu_{B}$)
and 25{\%} Mn$^{4+}$ ions ($\mu_{eff}$ $\approx$ 3.8 $ $ $\mu_{B}$)
in the system, resulting in an expected total effective moment
of $\mu_{eff}$ $\approx$ 4.6 $ $ $\mu_{B}$ which is consistent with our crude fitting analysis.

\begin{figure}[tbp]
\linespread{1}
\par
\begin{center}
\includegraphics[width=3.6in]{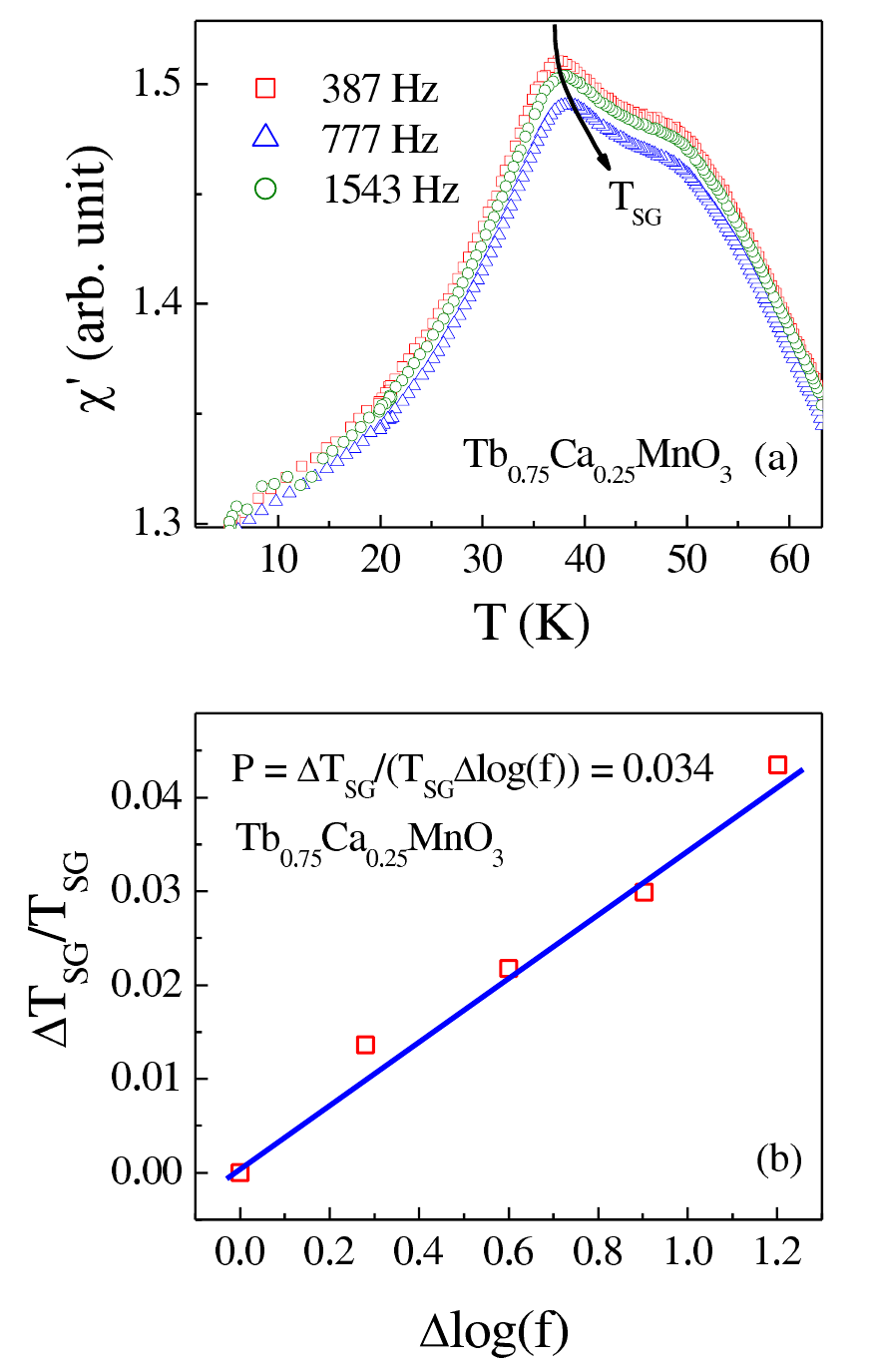}
\end{center}
\caption{(color online) (a) Temperature dependence of the real part of the AC susceptibility
for polycrystalline  Tb$_{0.75}$Ca$_{0.25}$MnO$_3$ under different frequencies.
(b) The frequency dependence of $\Delta T_{SG}$/$T_{SG}$.}
\end{figure}

The AC susceptibility was measured for Tb$_{0.75}$Ca$_{0.25}$MnO$_3$ to study the nature of
the transition at $T_{SG}$. As shown in Fig.~4(a), around $T_{SG}$ the AC susceptibility
presents
a frequency dependent peak. The Mydosh parameter $\Delta T \rm_{SG}$/[$T\rm_{SG}\Delta \rm{log}{(f)}$],
a quantitative measure of the frequency shift, is estimated to be 0.034 (Fig.~4(b)). This is of the
same order as the expected range of 0.004-0.018 for conventional spin glass systems.\cite{Mydosh}
The ZFC and FC splitting feature is also a characteristic behavior of a spin glass transition.

\begin{figure}[tbp]
\linespread{1}
\par
\begin{center}
\includegraphics[width=3.2in, trim=3 3 3 3, clip]{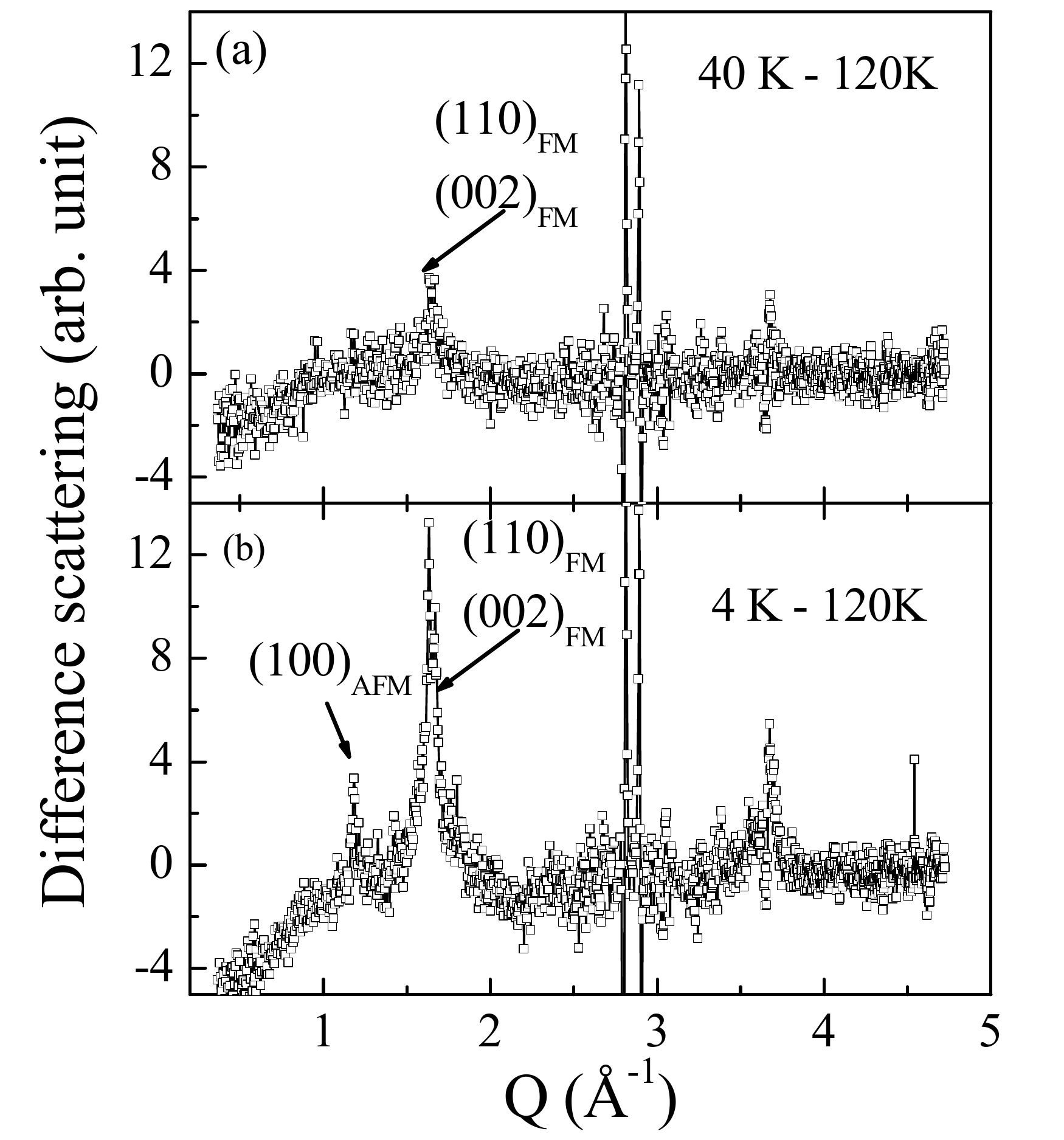}
\end{center}
\par
\caption{The differential scattering pattern for polycrystalline Tb$_{0.75}$Ca$_{0.25}$MnO$_3$
(a) between 40 K and 120 K and (b) between 4 K and 120 K.}
\end{figure}

The powder neutron diffraction measurements were performed on Tb$_{0.75}$Ca$_{0.25}$MnO$_3$, and the differential scattering results are shown in Fig. 5. The large feature near $Q = 2.8$ {\AA}$^{-1}$ is due to the change in lattice parameters. The negative differential signal at $Q$ $<$ 0.5 {\AA}$^{-1}$ is caused by the reduction in the paramagnetic scattering that exists above the ordering temperature. Note that the paramagnetic scattering follows the $Q$ dependence of the Mn$^{3+}$/Mn$^{4+}$ magnetic form factor. The observed magnetic peak at $Q=1.6$ {\AA}$^{-1}$ is broader than the instrument resolution. Its position suggests it is due to scattering from a short range ferromagnetic ordering that contributes to the (110) or (002) Bragg peaks which can be explained by a ferromagnetic ordering along either the $a$ or $b$ axes. This is consistent with the DC susceptibility
results showing the development of ferromagnetism below $T_1$. Figure 5(b) contains the
differential scattering between 4 K and 120 K. The data shows more pronounced scattering
that adds to the (110) or (002) peaks. In addition, there is another new magnetic peak
at $Q=1.2$ {\AA}$^{-1}$, matching the  (100) peak position. The presence of this lattice
forbidden (100) peak should be associated  with the development of an antiferromagnetic ordering. It is noteworthy that even at 4 K, these peaks are still
broader than the instrument resolution, and the correlation length derived from the (100) Lorentzian full peak width at half maximum is approximately $\xi$ = 50 {\AA}.

Therefore, it is concluded that Tb$_{0.75}$Ca$_{0.25}$MnO$_3$ develops ferromagnetic characteristics
below $T_1$ and enters a short-range ordered state below $T_{SG}$, but the latter contains
antiferromagnetic tendencies. Based on the similarity among the DC susceptibility measurements
of all the $R_{0.75}$Ca$_{0.25}$MnO$_3$ samples, this development of magnetism in the Tb sample
should occur in all the other samples as well.

\subsection{Single crystal Tb$_{0.75}$Ca$_{0.25}$MnO$_3$}

\begin{figure}[tbp]
\linespread{1}
\par
\begin{center}
\includegraphics[width=3.2in]{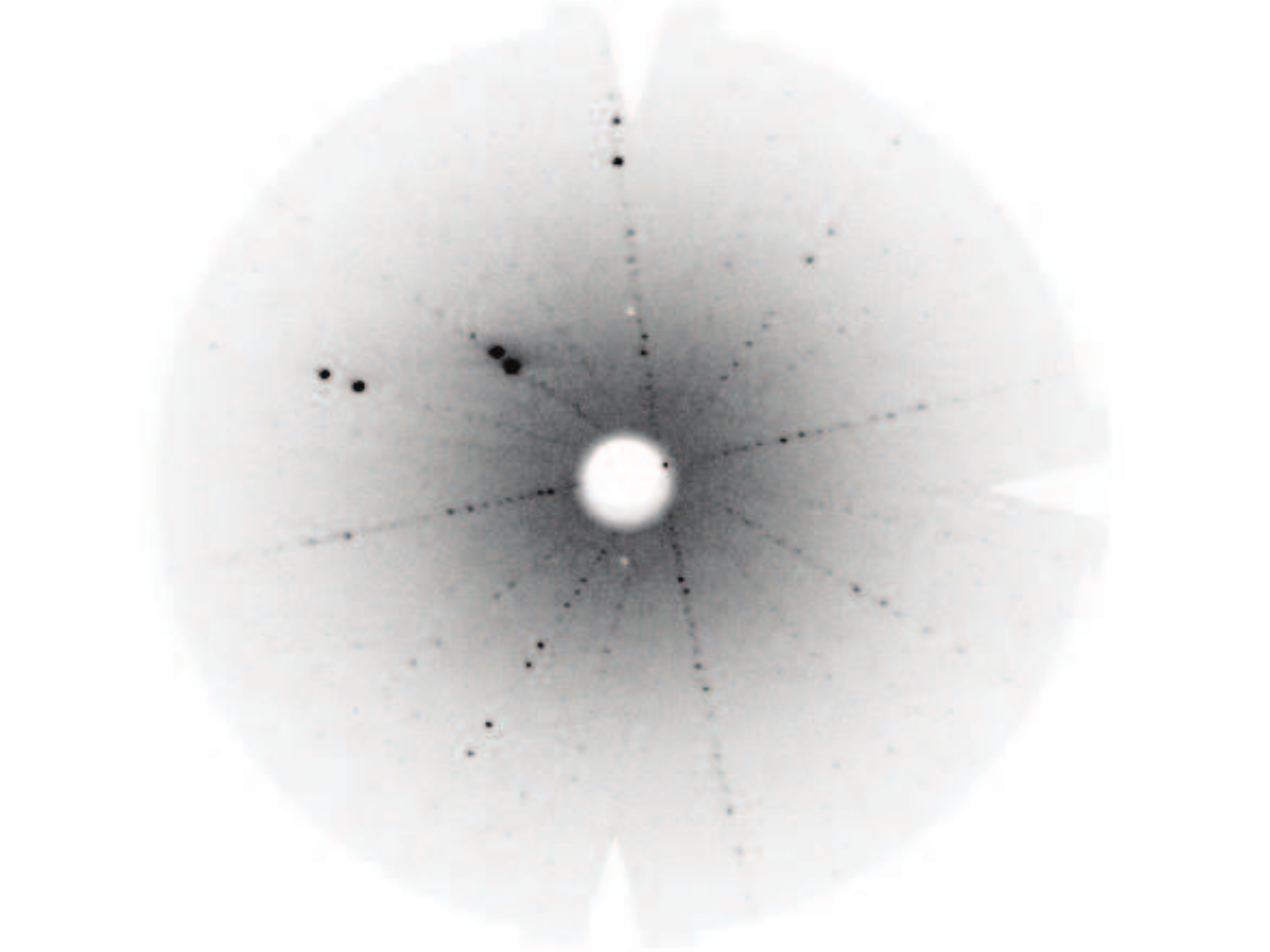}
\end{center}
\par
\caption{The Laue pattern oriented along the $[$001$]$ direction for the
grown Tb$_{0.75}$Ca$_{0.25}$MnO$_3$ single crystal.}
\end{figure}

In order to further clarify the nature of the magnetic properties of $R_{0.75}$Ca$_{0.25}$MnO$_3$,
we tried to grow single crystals for more detailed studies. We successfully grew
single crystals of Tb$_{0.75}$Ca$_{0.25}$MnO$_3$ by using the floating zone technique.
The obtained crystals cleave easily into several millimeter-long needle pieces.
The attempts to grow other $R_{0.75}$Ca$_{0.25}$MnO$_3$ samples all failed. Notably,
after melting at high temperatures, the Ho, Y, and Er samples show phase separation
by introducing the hexagonal phase into the orthorhombic phase. This fact suggests
that the orthorhombic phase of the Ho, Y, and Er samples is a meta-stable phase at
low temperatures. A Laue pattern measured on the grown Tb$_{0.75}$Ca$_{0.25}$MnO$_3$
single crystal is presented in Fig. 6.

\begin{figure}[tbp]
\linespread{1}
\par
\begin{center}
\includegraphics[width=2.5in]{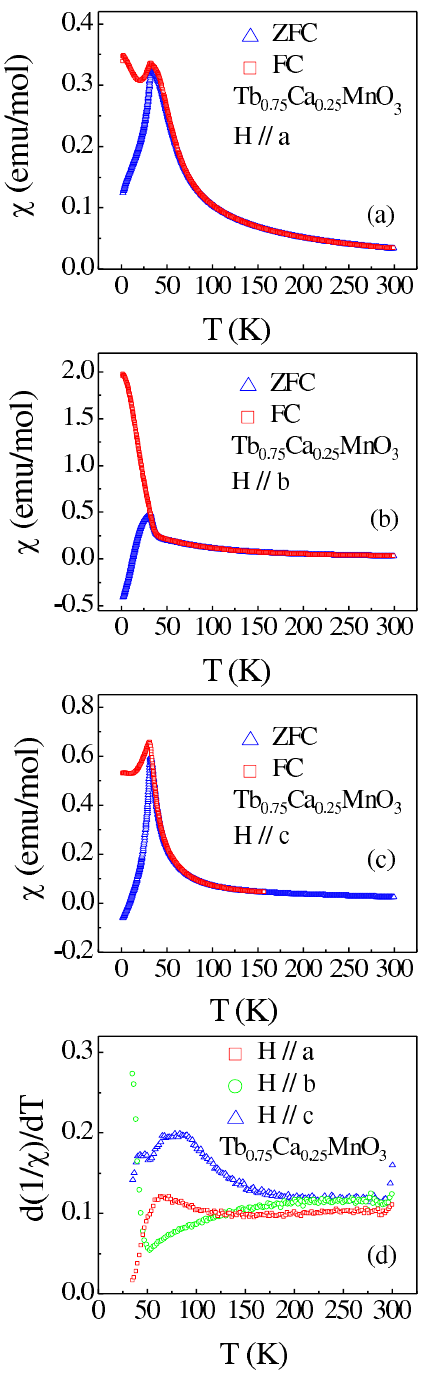}
\end{center}
\par
\caption{(color online) Temperature dependence of the DC susceptibility for a
single crystal of Tb$_{0.75}$Ca$_{0.25}$MnO$_3$ with (a) $H||a$, (b) $H||b$,
and (c) $H||c$. (d) Derivative of 1/$\chi$ with respect to temperature, with $H$ along the three axes.}
\end{figure}

Figure 7 shows the DC susceptibility of the Tb$_{0.75}$Ca$_{0.25}$MnO$_3$ single crystal with an
applied magnetic field $H$ along different axes. For $H||a$ and $H||c$, the results for $\chi$
are similar between the two data sets which shows a slope change around 80 K, as defined
by the peak obtained from the derivative of 1/$\chi$, and a ZFC and FC splitting around 30 K.
These features are consistent with the polycrystalline results.  For $H||b$, the $\chi$ results
are different from those corresponding to $H||a$ and $H||c$. First, the sharp increase
of $\chi$ now shifts to around 50 K. Second, the FC curve keeps increasing below 30 K
while the ZFC curve actually shows negative values which strongly suggests the existence
of ferromagnetic domains.

\begin{figure}[tbp]
\linespread{1}
\par
\begin{center}
\includegraphics[width=3.6in]{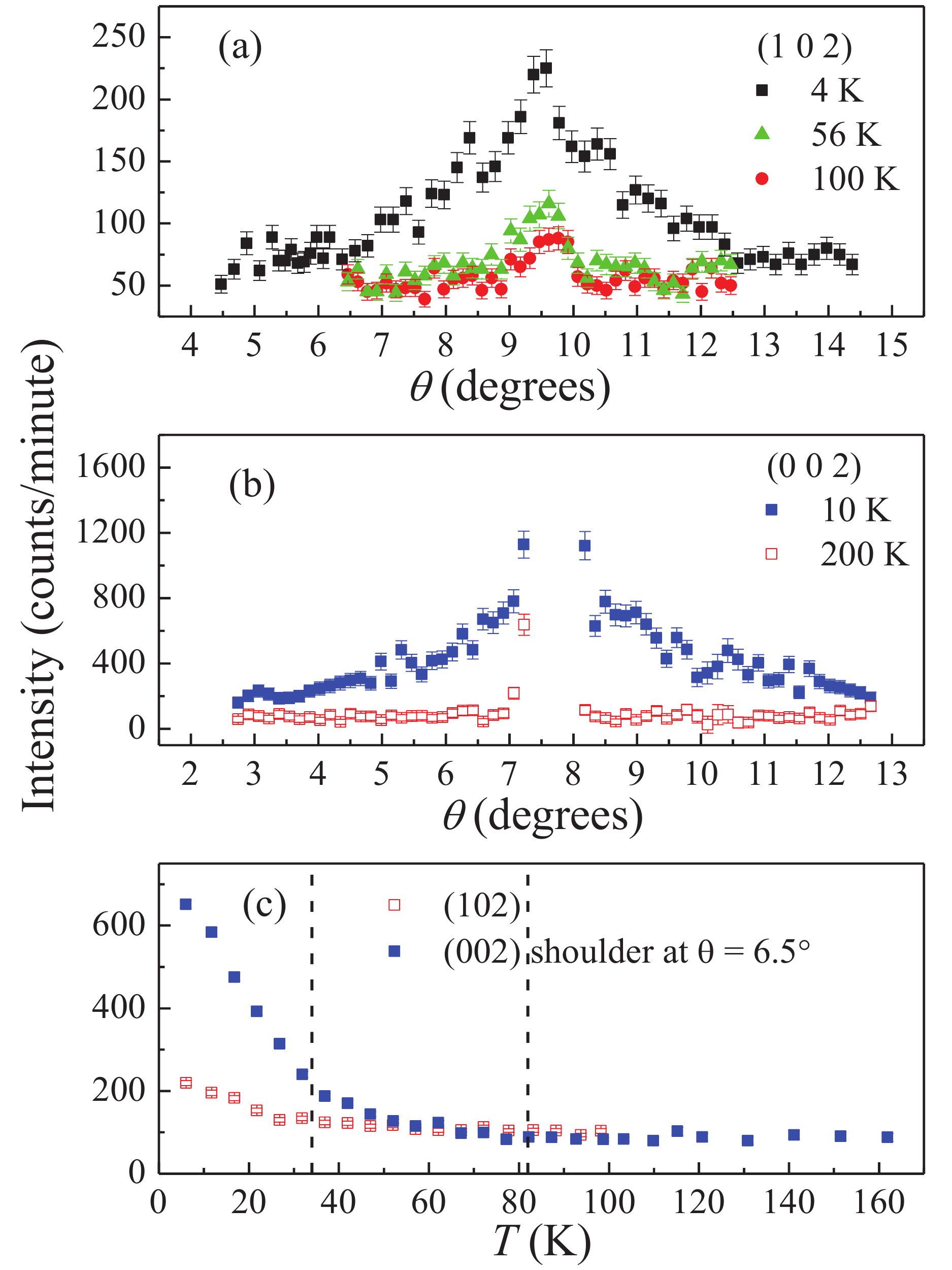}
\end{center}
\par
\caption{(color online) The (a) (102) and (b) (002) peaks at different temperatures for the
Tb$_{0.75}$Ca$_{0.25}$MnO$_3$ single crystal. (c) Temperature dependence of the (102) peak intensity and the (102) peak shoulder ($\theta$ = 6.5\degree) intensity. }
\end{figure}

\begin{table*}[tbp]
\par
\caption{Structural parameters for the $R_{0.75}$Ca$_{0.25}$MnO$_3$ samples ($R$ = Y, Tb, Dy, Ho, and Er) at room temperature (space group $Pbnm$) determined from refined XRD measurements. }
\label{t1}
\setlength{\tabcolsep}{0.55cm}
\begin{tabular}{ccccccc}
\hline
\hline
Refinement & Atom & Site & {\it x} & {\it y} & {\it z} & Occupancy \\ \hline
\multirow{5}{*}{\begin{tabular}[c]{@{}c@{}}XRD\\ $R$ = Y\\ $\chi^2$ = 0.976\\ (a)\end{tabular}} & Y & 4c &$   $ -0.01519(29) & 0.07107(15) & 1/4 & 0.37365(97) \\
 & Ca & 4c &$   $ -0.01519(29) & 0.07107(15) & 1/4 & 0.12635(97) \\
 & Mn & 4b &$   $ 1/2 & 0 & 0 & 0.50 \\
 & O1 & 4c &$   $ 0.11811(86) & 0.45828(96) & 1/4 & 0.50 \\
 & O2 & 8d &$   $ 0.69571(76) & 0.29496(73) & 0.04660(71) & 1.00 \\
\multicolumn{1}{l}{} & \multicolumn{1}{l}{} & \multicolumn{1}{l}{} & \multicolumn{1}{l}{} & \multicolumn{1}{l}{} & \multicolumn{1}{l}{} & \multicolumn{1}{l}{} \\
 & \multicolumn{6}{c}{$a$ = 5.290982(47), $b$ = 5.626886(54), $c$ = 7.448109(62)} \\
\multicolumn{1}{l}{} & \multicolumn{6}{l}{} \\
\multicolumn{1}{l}{} & \multicolumn{6}{c}{Overall B-factor = 1.9561} \\ \hline
\multirow{5}{*}{\begin{tabular}[c]{@{}c@{}}XRD\\ $R$ = Tb\\ $\chi^2$ = 0.332\\ (b)\end{tabular}} & Tb & 4c &$   $ -0.01760(61) & 0.06673(33) & 1/4 & 0.37403(133) \\
 & Ca & 4c &$   $ -0.01760(61) & 0.06673(33) & 1/4 & 0.12597(133) \\
 & Mn & 4b &$   $ 1/2 & 0 & 0 & 0.50 \\
 & O1 & 4c &$   $ 0.09883(233) & 0.46518(266) & 1/4 & 0.50 \\
 & O2 & 8d &$   $ 0.70392(241) & 0.29343(216) & 0.04061(225) & 1.00 \\
\multicolumn{1}{l}{} & \multicolumn{6}{l}{} \\
 & \multicolumn{6}{c}{$a$ = 5.333127(126), $b$ = 5.628304(138), $c$ = 7.493484(167)} \\
\multicolumn{1}{l}{} & \multicolumn{6}{l}{} \\
\multicolumn{1}{l}{} & \multicolumn{6}{c}{Overall B-factor = 2.4558} \\ \hline
\multirow{5}{*}{\begin{tabular}[c]{@{}c@{}}XRD\\ $R$ = Dy\\ $\chi^2$ = 0.697\\ (c)\end{tabular}} & Dy & 4c &$   $ -0.01432(49) & 0.06895(24) & 1/4 & 0.34640(94) \\
 & Ca & 4c &$   $ -0.01432(49) & 0.06895(24) & 1/4 & 0.15360(94) \\
 & Mn & 4b &$   $ 1/2 & 0 & 0 & 0.50 \\
 & O1 & 4c &$   $ 0.09971(160) & 0.47354(181) & 1/4 & 0.50 \\
 & O2 & 8d &$   $ 0.69111(145) & 0.30316(136) & 0.04818(147) & 1.00 \\
\multicolumn{1}{l}{} & \multicolumn{6}{l}{} \\
 & \multicolumn{6}{c}{$a$ = 5.317957(89), $b$ = 5.624245(103), $c$ = 7.482448(124)} \\
\multicolumn{1}{l}{} & \multicolumn{6}{l}{} \\
\multicolumn{1}{l}{} & \multicolumn{6}{c}{Overall B-factor = 2.3016} \\ \hline
\multirow{5}{*}{\begin{tabular}[c]{@{}c@{}}XRD\\ $R$ = Ho\\ $\chi^2$ = 0.628\\ (d)\end{tabular}} & Ho & 4c &$   $ -0.01513(31) & 0.06859(16) & 1/4 & 0.37244(69) \\
 & Ca & 4c &$   $ -0.01513(31) & 0.06859(16) & 1/4 & 0.12756(69) \\
 & Mn & 4b &$   $ 1/4 & 0 & 0 & 0.50 \\
 & O1 & 4c &$   $ 0.10984(114) & 0.48981(127) & 1/4 & 0.50 \\
 & O2 & 8d &$   $ 0.70930(115) & 0.31137(87) & 0.05731(90) & 1.00 \\
\multicolumn{1}{l}{} & \multicolumn{6}{l}{} \\
 & \multicolumn{6}{c}{$a$ = 5.291904(53), $b$ = 5.640843(58), $c$ = 7.449123(70)} \\
\multicolumn{1}{l}{} & \multicolumn{6}{l}{} \\
\multicolumn{1}{l}{} & \multicolumn{6}{c}{Overall B-factor = 2.2878} \\ \hline
\multirow{5}{*}{\begin{tabular}[c]{@{}c@{}}XRD\\ $R$ = Er\\ $\chi^2$ = 0.616\\ (e)\end{tabular}} & Er & 4c &$   $ -0.01729(20) & 0.07180(12) & 1/4 & 0.37492(51) \\
 & Ca & 4c &$   $ -0.01729(20) & 0.07180(12) & 1/4 & 0.12508(51) \\
 & Mn & 4b &$   $ 1/4 & 0 & 0 & 0.50 \\
 & O1 & 4c &$   $ 0.10613(96) & 0.46861(103) & 1/4 & 0.50 \\
 & O2 & 8d &$   $ 0.71026(92) & 0.31032(74) & 0.05153(64) & 1.00 \\
\multicolumn{1}{l}{} & \multicolumn{6}{l}{} \\
\multicolumn{1}{l}{} & \multicolumn{6}{c}{$a$ = 5.269618(41), $b$ = 5.647665(42), $c$ = 7.428105(54)} \\
\multicolumn{1}{l}{} & \multicolumn{6}{l}{} \\
\multicolumn{1}{l}{} & \multicolumn{6}{c}{Overall B-factor = 1.8764} \\
\hline
\hline
\end{tabular}

\end{table*}

\begin{figure}[tbp]
\linespread{1}
\par
\begin{center}
\includegraphics[width=\columnwidth]{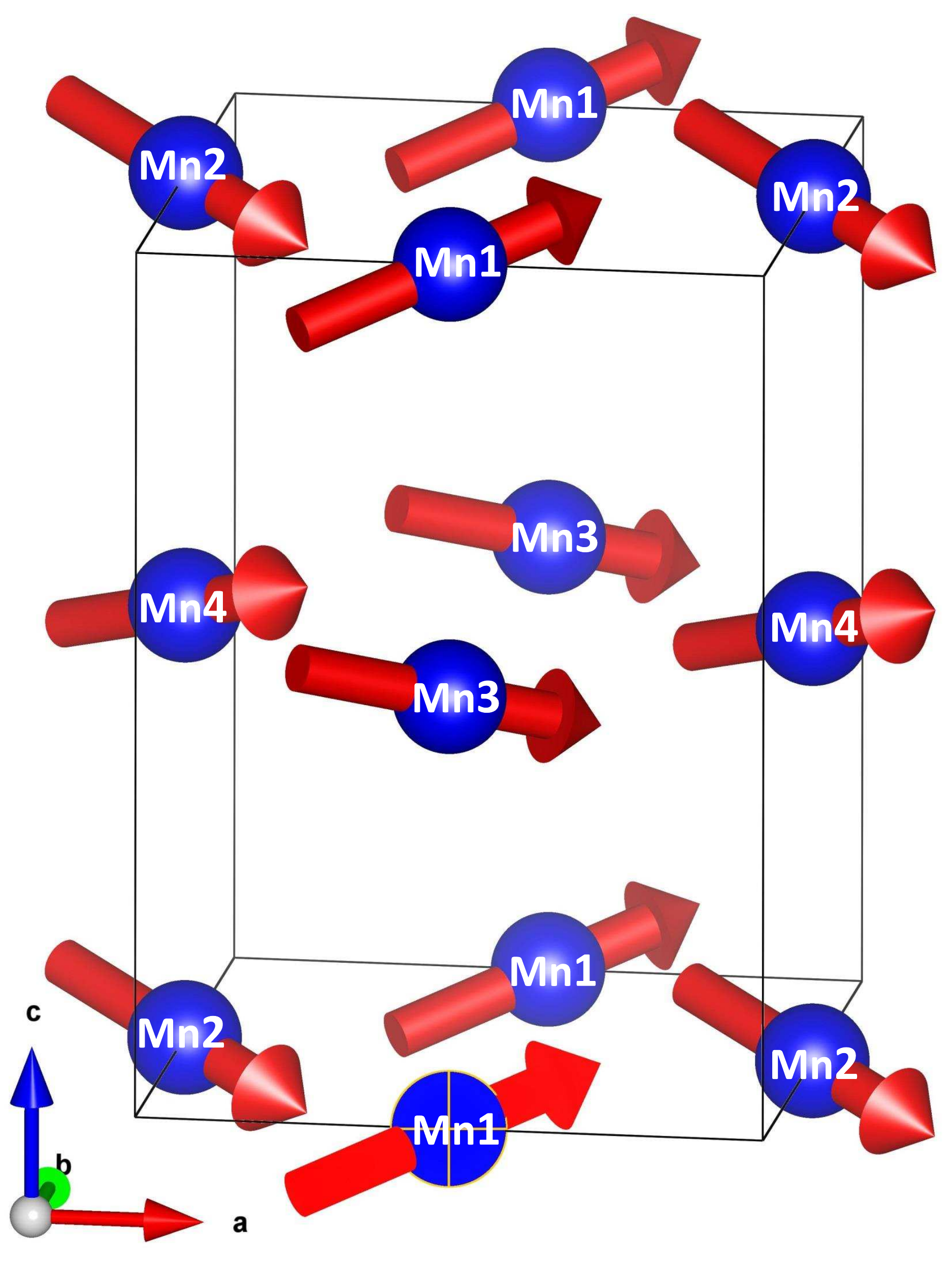}
\end{center}
\par
\caption{The novel spin state reported for the Tb$_{0.75}$Ca$_{0.25}$MnO$_3$ lattice.
The manganese sites Mn1, Mn2, Mn3, and Mn4 of Table II are indicated.
There are two canting angles between Mn's nearest-neighbor spins: an in-plane angle of $\sim77^\circ$ and an out-of-plane angle of $\sim32^\circ$.}
\end{figure}

As shown in Fig. 8, the single crystal neutron diffraction measurements on Tb$_{0.75}$Ca$_{0.25}$MnO$_3$
display two magnetic signals at low temperatures: (i) a weak broad peak around the
antiferromagnetic Bragg position  (102); (ii) a strong broad peak around the ferromagnetic
Bragg position (002). The temperature dependence of the intensity of the (102) peak shows
that it develops below 35 K, which is around $T_{SG}$. Meanwhile, the intensity of the shoulder
of the (002) peak at $\theta=6.5\degree$ starts developing below 80~K ($T_1$) and sharply
increases below 35~K. These features are consistent with the powder neutron diffraction
results. The refinement of this single crystal neutron diffraction data leads to the
identification of the magnetic ordering for temperatures lower than 35~K.

The details
of the spin structure are shown in Fig. 9. Here, we see that the Mn spins are canted both
in and out of the $a$-$b$ plane with canting angles of $\sim$77$^\circ$ and $\sim$32$^\circ$, respectively. The total refined moment of the system was 3.689 $\mu_{B}$,
as shown in Table II, smaller than our crude Curie-Weiss fit but still robust. This spin structure
is different from  the canonical A-type, E-type, CE-type, C-type, G-type, and spiral-type
antiferromagnetic phases observed in manganites before (using the canonical notation\cite{Dag1,Dag2}). Although the single crystal neutron data enables us to resolve the antiferromagnetic spin structure,
the broadness of the magnetic Bragg peaks strongly suggests that this antiferromagnetic ordering either has
short range characteristics or it has a clustered nature, as opposed to a true
long range magnetic ordering.

\section{DISCUSSION}

Only a few previous studies of narrow bandwidth $R_{1-x}$Ca$_x$MnO$_3$ have addressed
the magnetic properties of $R_{0.75}$Ca$_{0.25}$MnO$_3$.\cite{RCaJPhys, SOSQdopedPRL}
For example, Blasco {\it et al.} analyzed Tb$_{1-x}$Ca$_x$MnO$_3$ and they showed
that with increasing Ca doping the ferromagnetic interactions are enhanced while the
antiferromagnetic ordering is suppressed, and in particular the $x = 0.25$ sample
has a spin glass ground state.\cite{TbCaPRB1} Pena {\it et al.} focused
on Dy$_{1-x}$Ca$_x$MnO$_3$ and they also showed an enhanced ferromagnetic interaction
around 80~K for the $x = 0.25$ sample.\cite{DyCaSSS} In addition, a neutron powder
diffraction study of Y$_{0.7}$Ca$_{0.3}$MnO$_3$, which has a similar composition
as the Y$_{0.75}$Ca$_{0.25}$MnO$_3$ case studied here, shows that the magnetic ground
state has short range ordering with an antiferromagnetic nature due to the observed
broad Bragg magnetic peak around the (001) reflection.\cite{YCaAPA} Our reported data
here is consistent with all of these previous results. More importantly, our
systematic studies of $R_{0.75}$Ca$_{0.25}$MnO$_3$ point out the presence of
enhanced ferromagnetic tendencies around 80 K ($T_1$) and also the existence of
spin glass behavior around 30 K ($T_{SG}$). Moreover, these are general
behaviors for all $R_{0.75}$Ca$_{0.25}$MnO$_3$ with $R$ = Tb, Dy, Ho, Y, and Er.
With increasing $R$ ionic size, both $T_1$ and $T_{SG}$ generally increase;
in addition, as discussed below the spin glass ground state is compatible
with a short range ordering version of
a novel canted ferromagnetic spin state that was unveiled
based on our detailed neutron diffraction studies on Tb$_{0.75}$Ca$_{0.25}$MnO$_3$ (see Fig.~9).

\begin{figure}[tbp]
\linespread{1}
\par
\begin{center}
\includegraphics[width=3.8in]{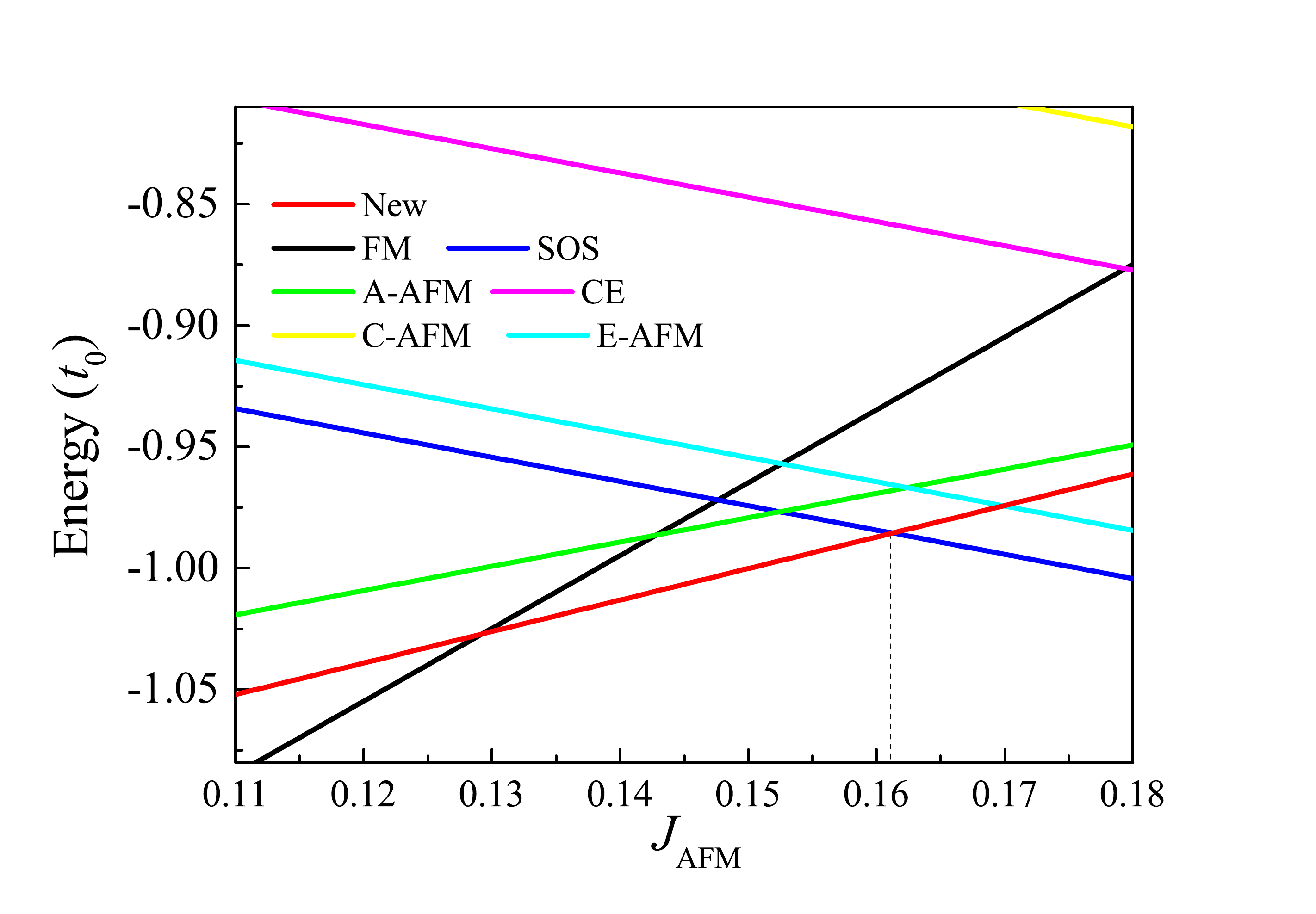}
\end{center}
\par
\caption{(color online) Energies (per site) of the magnetic states considered here,
as a function of $J_{\rm AFM}$. The energy unit is $t_0$. FM: ferromagnetic; A-AFM: A-type antiferromagnetic; C-AFM: C-type antiferromagnetic; E-AFM: E-type antiferromagnetic. These antiferromagnetic
as well as the CE labels are standard notations
in manganites.\cite{Dag1, Dag2} The SOS state is from Ref.~\onlinecite{SOSQdopedPRL}.
The novel canted state is denoted by ``New''.}
\end{figure}

To verify the exotic magnetic pattern obtained in our neutron analysis,
here a microscopic theoretical study is performed based on the standard
two-orbital double-exchange model.\cite{Dag1,Dag2,SSC} In the past decade,
this model Hamiltonian has been widely used to investigate a plethora of magnetic
phases and their associated physical characteristics, such as
colossal magnetoresistance and multiferroicity, in perovskite manganites.\cite{Dag1,Dag2}
The clear success of the previous efforts in this context
allows us to investigate with confidence the possibility of new phases in previously
unexplored regions of the phase diagrams via the double exchange model.
More explicitly, the model Hamiltonian used here reads as:
\begin{equation}
H=-‎‎\sum_{<ij>}^{\alpha\beta} t^{\bf r}_{\alpha\beta}
(\Omega_{ij} c^{\dagger}_{i,\alpha}c_{j,\beta}+h.c.)
+J_{\rm AFM}\sum_{<ij>}{{{\bf S}_i}\cdot{{\bf S}_j}}.
\end{equation}

This Hamiltonian contains two terms. The first term denotes
the standard two-orbital double-exchange hopping process for the
$e_g$ electrons between nearest-neighbor sites $i$ and $j$ of a three dimensional
cubic lattice for the manganese ions.
The operators $c^{\dagger}_{i,\alpha}$ ($c_{j,\beta}$) create (annihilate)
an $e_g$ electron at the orbital $\alpha$ ($\beta$) of the lattice site $i$ ($j$).
Working within the standard infinite Hund coupling approximation, shown to be qualitatively
correct for manganites,\cite{Dag1,Dag2} the spin of the $e_g$ electrons is always parallel
to the spin of the localized $t_{2g}$ degrees of freedom, ${\bf S}$,
generating the Berry phase:
$\Omega_{ij}=\cos(\theta_i/2)\cos(\theta_j/2)+\sin(\theta_i/2)\sin(\theta_j/2)\exp[-i(\phi_i-\phi_j)]$,
where $\theta$ and $\phi$ are the polar and azimuthal angles of the classical
$t_{2g}$  spins, respectively.\cite{Dag1, Dag2} The three nearest-neighbor (NN) hopping directions
are denoted by ${\bf r}$. Two $e_g$  orbitals ($a$: $x^2-y^2$ and $b$: $3z^2-r^2$)
are involved in the double-exchange process for manganites, with the hopping amplitudes along the
three axes given by:
\begin{eqnarray}
\nonumber t^x&=&\left(
\begin{array}{cc}
t^x_{aa} &  t^x_{ab} \\
t^x_{ba} &  t^x_{bb}
\end{array}
\right) =\frac{t_0}{4}\left(
\begin{array}{cc}
3 &  -\sqrt{3} \\
-\sqrt{3} &  1
\end{array}
\right),\\
\nonumber t^y&=&\left(
\begin{array}{cc}
t^y_{aa} &  t^y_{ab} \\
t^y_{ba} &  t^y_{bb}
\end{array}
\right) =\frac{t_0}{4}\left(
\begin{array}{cc}
3 &  \sqrt{3} \\
\sqrt{3} &  1
\end{array}
\right),\\
t^z&=&\left(
\begin{array}{cc}
t^z_{aa} &  t^z_{ab} \\
t^z_{ba} &  t^z_{bb}
\end{array}
\right) =t_0\left(
\begin{array}{cc}
0 &  0 \\
0 &  1
\end{array}
\right).
\end{eqnarray}

In our calculations, the hopping amplitude $t_0$ will be considered as the unit of energy.
This hopping can be roughly estimated to be $0.5$ eV.\cite{Dag1,Dag2}
The second term of the Hamiltonian is the antiferromagnetic superexchange
interaction between the NN $t_{2g}$ spins.

\begin{table*}[tbp]
\par
\caption{Magnetic moments for the single crystal Tb$_{0.75}$Ca$_{0.25}$MnO$_3$ sample at 4.2 K
(magnetic space group $Pbn'm'$) determined from refined neutron diffraction measurements.}
\label{t2}
\setlength{\tabcolsep}{0.58cm}
\begin{tabular}{cccccccc}
\hline
\hline
 & x & y & z & M$_x$ ($\mu_B$) & M$_y$ ($\mu_B$) & M$_z$ ($\mu_B$) & M ($\mu_B$) \\ \hline
Mn1 & 1/2 & 0 & 0   & 2.886( 62) &  2.063( 75) &  1.011(153) &   3.6888( 772)  \\
Mn2 & 0 & 1/2 & 0   & 2.886( 62) & -2.063( 75) & -1.011(153) &   3.6888( 772)  \\
Mn3 & 1/2 & 0 & 1/2 & 2.886( 62) &  2.063( 75) & -1.011(153) &   3.6888( 772)  \\
Mn4 & 0 & 1/2 & 1/2 & 2.886( 62) & -2.063( 75) &  1.011(153) &   3.6888( 772)  \\
\hline
\hline
\end{tabular}
\end{table*}

The typical value of the superexchange coupling $J_{\rm AFM}$ in manganites is approximately
$0.1t_{0}$ for the more widely studied manganites, such as La$_{1-x}$Sr$_{x}$MnO$_{3}$
and La$_{1-x}$Ca$_{x}$MnO$_{3}$, based on a variety of previous investigations.\cite{Dag1,Dag2}
The model studied here does not include the electron-lattice coupling, i.e.
the Jahn-Teller distortions, but this coupling can be partially taken into account by increasing
the superexchange strength. Our simplified model is expected to capture the main physics
of manganites and, thus, can generate a phase diagram qualitatively similar to the experimental
observations.\cite{Dag3} Therefore, it is acceptable to study, at least qualitatively, the properties of
the manganites discussed in the present publication using this simplified model.

For the quarter-doped manganites, previous investigations predicted a spin-orthogonal
stripe SOS phase in the very narrow bandwidth region ($J_{\rm AFM} > 0.17t_0$).\cite{SOSQdopedPRL}
The well known ferromagnetic phase observed in normal manganites appears in the opposite
side ($J_{\rm AFM} < 0.13t_0$). In the middle region, the Monte Carlo simulation did not provide
an unambiguous answer at that time because of metastabilities in the Monte Carlo
time evolution that are typically indicative of complex magnetic patterns.\cite{SOSQdopedPRL}
Since the spin pattern obtained in the neutron
study discussed above is neither the SOS nor the normal ferromagnetic phases, therefore it is necessary
to recheck the possible existence of new phases by taking into account the state discovered
experimentally here.

The energies for various fixed magnetic patterns are calculated in momentum space using a fine three dimensional grid and the results are shown in Fig. 10. With increasing $J_{\rm AFM}$, the ground state evolves from the initial
ferromagnetic state to the final SOS state, in agreement with previous investigations.\cite{SOSQdopedPRL}
However, the most interesting new result is that in the middle region ($0.129 t_0 < J_{\rm AFM} < 0.161 t_0$),
the newly discovered canted spin order state displays the lowest energy among all of the candidates investigated here.

According to these results, the case of Tb$_{0.75}$Ca$_{0.25}$MnO$_3$ should fall into
this middle region. The new canted phase provides a bridge
between the ferromagnetic and the SOS phases since it contains both a ferromagnetic
and noncollinear components. Thus, the pure SOS phase should be expected to be stable only in even
narrower bandwidth manganites.

It is worth clarifying that this canted state does {\it not} induce
ferroelectricity if we apply the inverse Dzyaloshinskii-Moriya (DM) mechanism where lattice
distortions leading to ferroelectricity are generated by special noncollinear magnetic
states.\cite{Sergienko:Prb} In the present case the local displacements will compensate between
nearest neighbors, and the global ferroelectric polarization will cancel out. Although not ferroelectric, the noncoplanar spin texture of the state unveiled here is very novel, and it can give rise to an intrinsic anomalous Hall effect.\cite{Chen:Prb10,Nagaosa:Rmp}

\section{CONCLUSIONS}

In this publication, we report detailed experimental studies of
$R_{0.75}$Ca$_{0.25}$MnO$_3$ ($R$ = Y, Tb, Dy, Ho, and Er)  polycrystals,
and a
Tb$_{0.75}$Ca$_{0.25}$MnO$_3$ single crystal, with focus on their magnetic properties.
The amount of Ca used corresponds to the hole quarter-doped case of
the widely discussed manganite
multiferroic perovskites. In general, we have observed the presence of ferromagnetic and spin-glass
tendencies in all the samples studied. Our main discovery, using the Tb-based single crystal,
is that the spin-glass region
appears dominated by the short-range order of a new canted, and thus noncollinear, magnetic state.
The theoretical study of the double-exchange model presented here shows that
in a reduced region of parameter space the new state has indeed lower energy
than the two states previously believed to be dominant at
quarter doping in narrow bandwidth manganites, namely the FM and SOS states.
The results reported here illustrate that doped manganite multiferroic compounds harbor
magnetic states that are more complex than previously anticipated. Our present efforts
are expected to pave the way and motivate more detailed studies of these mainly unexplored
exotic materials that have potential for functional applications.

\begin{acknowledgments}
R. S., Z.L.D., and H.D.Z. thank the support from NSF-DMR through Award DMR-1350002.
The research at HFIR/ORNL was sponsored by the Scientific User Facilities Division
(H.B.C., O.V.G, J.M.), Office of Basic Energy Sciences, US Department of Energy.
S.D. was supported by National Natural Science Foundation of China (Grant Nos. 51322206).
E.D. was supported by the National Science Foundation under Grant No. DMR-1404375. The work at NHMFL is supported by Grant No. NSF-DMR-1157490 and the State of Florida and by the additional funding from NHMFL User Collaboration Support Grant.

\end{acknowledgments}

\end{document}